\documentclass[]{article}
\usepackage{amsmath}  % needed for \tfrac, \bmatrix, etc.
\usepackage{amsfonts} % needed for bold Greek, Fraktur, and blackboard bold
\usepackage{graphicx} % needed for figures
\usepackage{braket}
\usepackage{amssymb}
\usepackage{hyperref}
\usepackage{subcaption}
\usepackage{cleveref}
\usepackage{mathtools}
\usepackage{float}
\usepackage{tikz}
\usetikzlibrary{angles,patterns,calc}
\usepackage[toc,page]{appendix}
\usepackage{authblk}
\usepackage[backend=biber,style=alphabetic,sorting=ynt]{biblatex}
\title{Explaining Grover’s algorithm with a colony of ants: a pedagogical model for making quantum technology comprehensible}

\author{Merel A. Schalkers\footnote{Corresponding author: \url{m.a.schalkers@tudelft.nl}}}
\affil{Delft University of Technology, Department of Applied Mathematics}
\author{Kamiel Dankers}
\author{Michael Wimmer}
\affil{Qutech and Kavli Institute of Nanoscience, Delft University of Technology}
\author{Pieter Vermaas}
\affil{Delft University of Technology, Department of Philosophy}

\date{\today}

\begin{document}

\maketitle

\begin{abstract}
\noindent The rapid growth of quantum technologies requires an increasing number of physicists, computer scientists, and engineers who can work on these technologies. For educating these professionals, quantum mechanics should stop being perceived as incomprehensible. In this paper we contribute to this change by presenting a pedagogical model for explaining Grover’s search algorithm, a prominent quantum algorithm. This model visualizes the three main steps of Grover's algorithm and, in addition to explaining the algorithm itself, introduces three key principles of quantum mechanics: superposition, interference, and state collapse at measurement. The pedagogical model, visualized by a video, is called the \textit{Ant Colony Maze model}. It represents the search problems as finding the exit of a maze, and visualizes Grover’s search algorithm as a strategy by which a colony of ants finds that exit. 
\end{abstract}

\maketitle 

\section{Introduction}
\noindent The rapid growth of quantum technologies requires a substantially increasing number of physicists, computer scientists, and engineers working on these technologies. This development can be expected to remain in place for the coming decade, and necessitates educating large contingents of students who can work on quantum technologies. While being of a different nature than constructing quantum computers, quantum internet, or quantum sensors, scaling up teaching for future quantum technologists is urgent and challenging as well. Having the famous quote by Richard Feynman in mind -- \textit{I think I can safely say that nobody understands quantum mechanics}\cite{Feynman1995} -- quantum mechanics, being the basis of quantum technologies, has the connotation of being incomprehensible, framing understanding this basis as a quest only a few privileged can realize. In this paper we aim to contributing to changing this perspective on quantum mechanics and enabling the education of larger contingents of students. Specifically, we contribute to education in quantum technology by presenting a pedagogical model for explaining a quantum algorithm: Grover’s search algorithm \cite{Grover1996}.

We submit the \textit{Ant Colony Maze model} as a pedagogical model for explaining Grover’s algorithm in teaching and for introducing key concepts of quantum mechanics. Moreover, by the graphic nature of the model, it can also provide nonprofessionals an informed feel of how quantum computing works, thus also contributing to an understanding of quantum technologies in society at large \cite{Vermaas2017}, breaking away from the Feynmanian frame that understanding the quantum realm is a privilege of the few.

Section~\ref{sec_quantum_comp} introduces quantum computing and Section~\ref{sec_grover} describes Grover’s algorithm. In Section~\ref{sec_ants} we present our Ant Colony Maze model. Existing visualizations of Grover's algorithm are reviewed in  Section~\ref{sec_visual} and compared to our model in Section~\ref{sec_comparison}. Section~\ref{sec_conclusions} is for conclusions.

\section{Quantum computers}
\label{sec_quantum_comp}
\subsection{Overview}

\noindent Since its discovery in the 1920s quantum mechanics have played an important role in physics. Not long after, quantum mechanical phenomena were embraced by engineers, with the potential of exploiting rather than just describing its properties.

In 1980 the idea of quantum computation was first suggested by Paul Benioff \cite{Benioff1980}. Later Yuri Manin and Richard Feynman proposed it for efficiently simulating quantum processes \cite{Manin2007, Feynman1982}, Quantum computing remained a rather small area of research until in 1994 Peter Shor published an algorithm for efficiently factorizing large numbers with quantum computers, thus highlighting the field's potential.

Nowadays both governmental organizations and private companies have invested millions in the technology with private companies being in the lead in developing quantum computers. Google was the first to claim a \emph{quantum advantage}, i.e., a quantum computer solving a problem faster than a classical computer \cite{Google2019}. Moreover, IBM currently offering 24 quantum computers online, the largest machine being the IBM Ospray containing 433 physical quantum bits, or \textit{qubits} \cite{ibmquantum}. Other companies such as D-Wave, Rigetti, Honeywell, and Google, and academic research institutes such as Oxford University and Delft University of Technology, also offer quantum computers for running algorithms via the cloud. 
Despite the great progress that has been made in the last decades and specifically the last couple of years, there are still significant challenges to be overcome before quantum computing can live up to its promises. Before we can expect large scale industrial applications more qubits with a higher precision and stability will be required.\footnote{Experts often make a distinction between noisy physical qubits and theoretical noiseless qubits, called \textit{logical qubits}. Research focuses on composing these logical qubits of large sets of physical qubits, thus increasing the task to create quantum computers with large numbers of physical qubits.}

\subsection{Calculating with qubits}
\noindent Calculating with qubits differs from calculating with classical bits in two central ways, and precisely these differences provide the potential and subtleties of quantum computing. 

Just like classical bits, qubits can be in states that correspond to 0 or 1. These states are written as $\ket{0}$ and $\ket{1}$ in quantum computing. A first difference between classical bits and qubits is that the latter can also be in states that are so-called superpositions of 0 and 1. Such states $\ket{\phi}$ are a complex linear combination of the states $\ket{0}$ and $\ket{1}$ that corresponds to 0 and 1: 
\begin{equation}
    \ket{\phi} = \alpha_0\ket{0} + \alpha_1\ket{1},
\end{equation}
with $\alpha_0$ and $\alpha_1$ complex numbers in $\mathbb{C}$, and $|\alpha_0|^2 + |\alpha_1|^2 = 1$. 

For determining whether a qubit has a value 0 or 1, one has to perform a measurement on the qubit in the computational basis. \footnote{A quantum measurement always needs to be performed with respect to a certain basis. The computational basis is the standard choice consisting of the basis $\ket{0}$, $\ket{1}$ that corresponds to the qubit values 0 and 1.} If the qubit is initially in state $\ket{0}$, one finds 0 with the measurement, and when the qubit is in state $\ket{1}$, one finds 1. If however the state of the qubit is a superposition of $\ket{0}$ and $\ket{1}$, there is $|\alpha_0|^2$ probability of finding 0 and $|\alpha_1|^2$ probability of finding 1. The amplitudes $\alpha_0$ and $\alpha_1$ in the superposition thus represent the probabilities of finding 0 and 1 upon measurement on the qubit. 

When extending the idea of superposition to multiple qubits, the quantum state $\ket{\phi}$ of $n$ qubits is a combination of the states of the separate qubits. For $n$ qubits we have $2^n$ basis states (as each separate qubit has two basis states $\ket{0}$ and $\ket{1}$), meaning that we can see an $n$-qubit state as a complex combination of $N=2^n$ basis states $\ket{i}$. So $n$ qubits have a combined quantum state that spans a space that is exponentially large in the number of qubits.
It is, however, important to keep in mind that only one basis state $\ket{i}$ will be found upon measurement. Mathematically we can express a general $n$-qubit quantum state as:
\begin{equation}\label{eq:quantum_state}
    \ket{\phi} = \sum_{i=0}^{2^n-1} \alpha_i \ket{i}
\end{equation}
with $\alpha_i \in \mathbb{C}$ for all $i \in \{0,1,\dots,2^n-1\}$ and $\sum_{i=0}^{2^n-1} |\alpha_i|^2=1$. Now the probability of finding the state $\ket{i}$ upon measurement is $|\alpha_i|^2$. To translate such a state $\ket{i}$ back to the value of the separate qubits, we can express the number $i$ in binary, to see precisely which qubit has which value in this basis state.

When we start performing calculations on $n$ qubits, it is easiest to interpret the quantum state of these qubits as a large vector in complex space. More specifically, an $n$-qubit quantum state can be seen as a vector $\ket{\phi}$ in $\mathbb{C}^{2^n}$. We can then interpret an operation on such a quantum state as a multiplication of the state $\ket{\phi}$ by a matrix $U$ in $\mathbb{C}^{2^n \times 2^n}$. Furthermore, since we require that the amplitudes $\alpha_i$ of a quantum state as expressed in Equation \eqref{eq:quantum_state} have the property that their absolute values squared sum to 1, this matrix $U$ must be unitary. 

Having introduced the basic qubit properties and how to apply operations to them, it can intuitively be seen how quantum computing can lead to a computational speed-up when properly exploited.  Assume we have $n$ qubits. The quantum state $\ket{\phi}$ of these qubits can then be written as a linear combination of $r \leq 2^n$ linearly independent vectors $\ket{\phi_i} \in \mathbb{C}^{2^n}$, that is, as $\ket{\phi} = \sum_{i=1}^{r} \alpha_i \ket{\phi_i}$. Now, when applying the unitary operation $U$ to the quantum state $\ket{\phi}$ we get  $U\ket{\phi} = \sum_{i=1}^{r} \alpha_i U\ket{\phi_i}$ using basic linear algebra. This implies that quantum computing allows us to perform in parallel $r$ different operations on $r$ different states $\ket{\phi_i}$, with the number $r$ potentially growing exponentially with the number $n$ of qubits. This is often referred to as ``quantum parallelism". Notice, however, that since we will only find \emph{one} computational basis state upon measurement, this parallelism is not immediately useful. More precisely, a quantum algorithm needs to consist of quantum operations that yield a final state with a very high probability in the desired solution, such that it can be found in a measurement.

By exploiting these properties of quantum computers, quantum algorithms can be designed that solve problems potentially more efficient than algorithms for classical computers. 

\section{Grover's search algorithm} 
\label{sec_grover}
\noindent One quantum algorithm that achieves a speed-up over its classical counterparts is Grover's search algorithm. Grover's algorithm was first described in 1996 \cite{Grover1996}, and remains an important part of education in quantum technology to this date. This is due to the fact that it is a relatively simple algorithm that elegantly exploits the potential of quantum computing, thus being a good example for learning the tricks of the trade.  

Even though Grover's algorithm might be relatively simple, novice students may focus on understanding the mathematics and miss to see the steps in the algorithm that are enabled by quantum mechanics. This paper gives a pedagogical model for explaining these quantum steps in Grover's algorithm in an intuitive way that does not request a deep knowledge of the underlying mathematics. There are three such steps and we describe them as each exploiting a key principle of quantum mechanics. Later we use this to show how our pedagogical model does a better job at introducing these principles than existing models.

\subsection{The search problem}
\noindent The search problem tackled by Grover's algorithm is, in computer science terms, searching an unstructured database: the problem is to find items from a search space that have a certain property. For example, this could be items fulfilling a set of mathematical equations. This search problem can be translated into quantum language as the task of finding within a large space $\mathbb{C}^{2^n}$ spanned by $2^n$ basis states those basis states $\ket{i}$ that represent the items with the desired property. 

The search problem in Grover's algorithm is defined in terms of an \textit{oracle} function. Such a quantum oracle function can recognize the correct states $\ket{i}$ within the large set of all quantum basis states: when giving a general quantum state $\ket{\phi}$ as input, the oracle function recognizes the basis states $\ket{i}$ with the desired property and marks them \textcolor{black}{with a minus sign. It is important here to n}ote that the oracle does not actually `know' which are the basis states $\ket{i}$ representing items with the desired property; it can only recognize whether a basis state is a `good' one and which basis state is a `bad' one, when given an input. Generally, recognizing a correct solution is easier than knowing it -- for example it is easier to check if a solution fulfills an equation than to derive the solution for that equation. We denote this quantum oracle by $O_X$ where $X$ is the set of indices $i$ of basis states $\ket{i}$ that represent items with the desired property. \textcolor{black}{This quantum oracle is hence problem-dependent. As such the oracle function is the only part of the quantum circuit of which the implementation will be different depending on problem, the remainder of the circuit of Grover's algorithm is universal.}

One could compare the problem of searching an unstructured database to finding in a maze a path that leads to the exit of the maze, and we use this comparison in our pedagogical model for explaining Grover's algorithm. Most paths in a maze do not lead to the exit, but some paths have this special property, and we want one of those. From this example the difference between knowing and recognizing a correct solution can be better understood: knowing which paths in a maze lead to the exit is much more difficult than simply recognizing whether or not a given path does so. 

\subsection{Description of Grover's algorithm}
\label{subsec_grover}
\noindent Grover's algorithm can be divided into three steps. The first can be interpreted as creating a state covering all items in the database. The second step as homing in on the `good' items with the oracle function. Subsequently the third step reveals one of those `good' items.
\subsubsection{Step 1: spreading out over the space}
\noindent We start with our qubits in the state $\ket{0}$, and the first thing we want to do is spread out our quantum state over the full space $\mathbb{C}^{2^n}$. We can do this by applying an operation that is standard in quantum computing: applying the Hadamard gate\footnote{$H=\frac{1}{\sqrt{2}}\begin{bmatrix} 1 & 1 \\ 1 & -1 \end{bmatrix} $} $H$ to all $n$ qubits. This operation gives us a new state, called $\ket{\mathcal{U}}$, which is: 
\begin{equation}\label{eq:equal superposition}
    \ket{\mathcal{U}}=H^{\otimes n}\ket{0} = \frac{1}{\sqrt{2^n}}\sum_{i=0}^{2^n-1} \ket{i}.
\end{equation}
This means that our qubits are brought in a state that is an equal superposition of all basis states and thus spread out over the full space $\mathbb{C}^{2^n}$. By being in this state $\ket{\mathcal{U}}$, we know with certainty that all the `good' basis state corresponding to a solution of our search problem are also present in the quantum state of the $n$ qubits. We can thus write this state $\ket{\mathcal{U}}$ as a linear combination of the `good' states and of the `bad' states. The superposition of the `good' states is $\ket{\mathcal{G}}=\frac{1}{\sqrt{\text{dim}(X)}}\sum_{i:i \in X}\ket{i}$, and the superposition of the `bad' states is $\ket{\mathcal{B}}=\frac{1}{\sqrt{2^n-\text{dim}(X)}}\sum_{i:i \notin X}\ket{i}$. Then we get: 
\begin{equation}\label{eq:grover_initial_state}
    \ket{\mathcal{U}}=\sin(\theta)\ket{\mathcal{G}} + \cos(\theta)\ket{\mathcal{B}},
\end{equation} 
with $\theta = \arcsin(\sqrt{\text{dim}(X)/2^n})$.

In this first step Grover's algorithm creates a state that is a linear combination of all basis states. In terms of our maze metaphor this means that after step 1 Grover gives a state that describes all possible paths in the maze: the `good' ones that lead to the exit, and the `bad' ones that do not. 

\subsubsection{Step 2: Grover iterate} \label{sec:grover_iterate}
\noindent After having spread out the quantum state to be in an equal superposition of all possible basis states, we use the oracle function to `mark' the `good' basis states that represent the items with the searched for desired property. Specifically the oracle $O_X$ will multiply the `good' basis states with the factor $-1$. This leads to the quantum state:
\begin{equation}
    \ket{\mathcal{U}_1} = O_X \ket{\mathcal{U}} = \frac{1}{\sqrt{n}}\sum_{i=0}^{2^n-1} \left ( -1 \right )^{x_i} \ket{i}.
\end{equation}
In the above equation $x_i=1$ if $i \in X$ and $x_i=0$ otherwise. This operation is equivalent to reflecting $\ket{\mathcal{U}}$ through the equal superposition $\ket{\mathcal{B}}$ of all `bad' states: $O_X = 2 \ket{\mathcal{B}} \bra{\mathcal{B}} - I$. 

Subsequently we apply another reflection operation $H^{\otimes n}R_0H^{\otimes n} =  2\ket{\mathcal{U}}\bra{\mathcal{U}}-I$ on the newly obtained state $O_X \ket{\mathcal{U}}$. Here, $R_0$ marks all but the $\ket{0}$ state with a factor $-1$ \textcolor{black}{and combined with the  $H^{\otimes n}$ operation, which turns the $\ket{0}$ state into $\ket{\mathcal{U}}$, it can be seen that} the total operation is a reflection through the state $\ket{\mathcal{U}}$. Both reflections are together called a \textit{Grover iterate}, and after one such an iterate, we are in the state \cite{DeWolf}:
\begin{equation}\label{eq:grover_intermediate_state}
\ket{\mathcal{U}_2} = \sin(2\theta + \theta)\ket{\mathcal{G}} + \cos(2\theta + \theta)\ket{\mathcal{B}}. 
\end{equation}
\textcolor{black}{The combination of the two reflections, first around the 'bad' states $\ket{B}$ and finally around the starting state $\ket{U}$ has the combined effect of rotating away from the starting state $\ket{\mathcal{U}}$ by an angle $2\theta$ such that the resulting state is closer to the `good' states. That the combination of the two reflections results in this rotation can be seen geometrically from Figure \ref{fig:arrow_diagram}.}

Applying this Grover iterate $k$ times brings us to the state:
\begin{equation}\label{eq:final superpos}
\ket{\mathcal{U}_{k}} = \sin((2k+1)\theta)\ket{\mathcal{G}} + \cos((2k+1)\theta)\ket{\mathcal{B}}.
\end{equation}
Using this expression it can be seen that if we chose $k$ as $\frac{\pi}{4\theta} - \frac{1}{2}$, we get $\sin((2k+1)\theta)=1$ and $\ket{\mathcal{U}_{k}} = \ket{\mathcal{G}}$. This choice is possible if the value $\frac{\pi}{4\theta} - \frac{1}{2}$ is an integer; if this is not the case, we take $k$ to be close to this value, and we get $\ket{\mathcal{U}} \approx \ket{\mathcal{G}}$. The idea is that we perform this rotation $k$ times such that the $n$ qubits are in a state that is very close or equal to the state $\ket{\mathcal{G}}$ that constitutes the equal superposition of the `good' states. In both cases we end up with a state that is very close or even equal to a superposition of the `good' states. 

Returning to the maze: Grover's algorithm brings the state of the $n$ qubits by step 2 to one where most of the weight is on the `good' paths that lead to the exit. Moreover, the operations of steps 1 and 2 can be carried out in a predetermined amount of time. 

\subsubsection{Step 3: Measurement}
\noindent Finally we measure the state of the $n$ qubits. After step 2 the state of the $n$ qubits is $\ket{\mathcal{U}_{k}}$, which is equal or very close to the superposition $\ket{\mathcal{G}}$ over all possible `good' basis states. 
Then, upon measurement, the probability of finding one of the `good' basis states that represent an item with the desired property, is equal, or almost equal, to one. Moreover, in quantum mechanics the state of a system collapses after the measurement to the basis state that corresponds to the found outcome. Hence, after the measurement the qubits are with probability 1, or almost 1, in a `good' state, that is, in a basis state that represents a solution to the search problem.

In terms of the maze: by the measurement in this last step we find one path. The more weight on a path, the larger the chance that we will get this path. And Grover's algorithm ensures that the `good' paths that lead to the exit have together a weight that is almost 1. So the chance that we find such a successful path becomes very large. 

\subsection{The three key quantum principles in Grover's algorithm}
\label{sec_three_steps}

\noindent Each of the three steps of Grover's algorithm makes use of one of the key quantum principles: superposition, interference, and measurement. Hence, explaining the algorithm is a good way to introducing students to the computations means that quantum mechanics offers.

Step 1 of Grover's algorithm introduces superposition as a key quantum principle: the initial state of the quantum bits in the quantum computer is brought in a linear combination of all basis states, where some basis state represents a possible solution within the search and has a certain amplitude. Students then learn that qubits can be in states that are superpositions of other states. 

Step 2 introduces the key principle of interference. The quantum state of the qubits is made to evolve in a way that the amplitudes of the basis states that represent `bad' solutions to the search problem become close to zero, and that the amplitudes of the `good' basis states grow. Students learn that states of qubits can interfere to fade out or become more present. 

Step 3 of Grover's algorithm presents the key quantum principle of measurement. The state of the qubits collapses with a probability close to 1 to a basis state representing a solution within the search. Students learn that at measurements the states of qubits make transitions from superpositions of basis states to single basis states.

\section{The Ant Colony Maze model}
\label{sec_ants}
\noindent In our description of Grover's algorithm in Section \ref{sec_grover} we already used the metaphor of the maze for Grovers algorithm. Now we introduce the Ant Colony Maze model as a pedagogical model for explaining the workings of Grover's search algorithm on a quantum computer. This model comes as a video that can be accessed online.\footnote{https://www.youtube.com/watch?v=eGYuTuTxojE 
} We first introduce the ingredients of the Ant Colony Maze model and how they relate to the elements of quantum computing. Then we explain how the dynamics in the Ant Colony Maze model represents the three steps of Grover's search algorithm. 

\subsection{The ingredients}
\noindent The Ant Colony Maze model consists of a maze with different paths, and a colony of ants trying to find the exit of the maze, marked by food (see Figure \ref{fig:screenshot1}, which is a still from the video representing the model). The ant colony as a whole represents the quantum computer. The paths in the maze represent the different possible solution states $\ket{i}$ of Equation \eqref{eq:equal superposition} and the path marked with food represents the correct outcome state $\ket{\mathcal{G}}$ for the given instance of Grover's search problem. Each ant represents an equal discrete density, the amount of ants on a certain path $i$ in the maze after $k$ timesteps thus represent the weight of the basis state $\ket{i}$ representing that path in the superposition of Equation \eqref{eq:final superpos}. This means that after $k$ timesteps the amount of ants on a path (see, e.g., the video still in Figure~\ref{fig:screenshot2}) represent a discrete probability of measuring that path; in the model it is assumed that upon a measurement of the ant colony, only one ant (each with equal probability) will be found, and that the path this ant is on represents the found state $\ket{i}$ upon measurement. The ants traveling on a path that leads to food will keep travelling on this path and the ants on an unsuccessful path will choose another path to try for food. This mechanism represents the role of the oracle, as it leads to an increase of the ants traveling on the path that leads to the exit of the maze in each timestep.

Below we explain how the ants move through the maze over time and how this behavior can be interpreted as the workings of a quantum computer running Grover's search algorithm as described in Section~\ref{sec_grover}. 

\subsection{The three steps}
\begin{figure}[ht!]
\begin{subfigure}{.47\textwidth}
  \centering
\includegraphics[width=\textwidth]{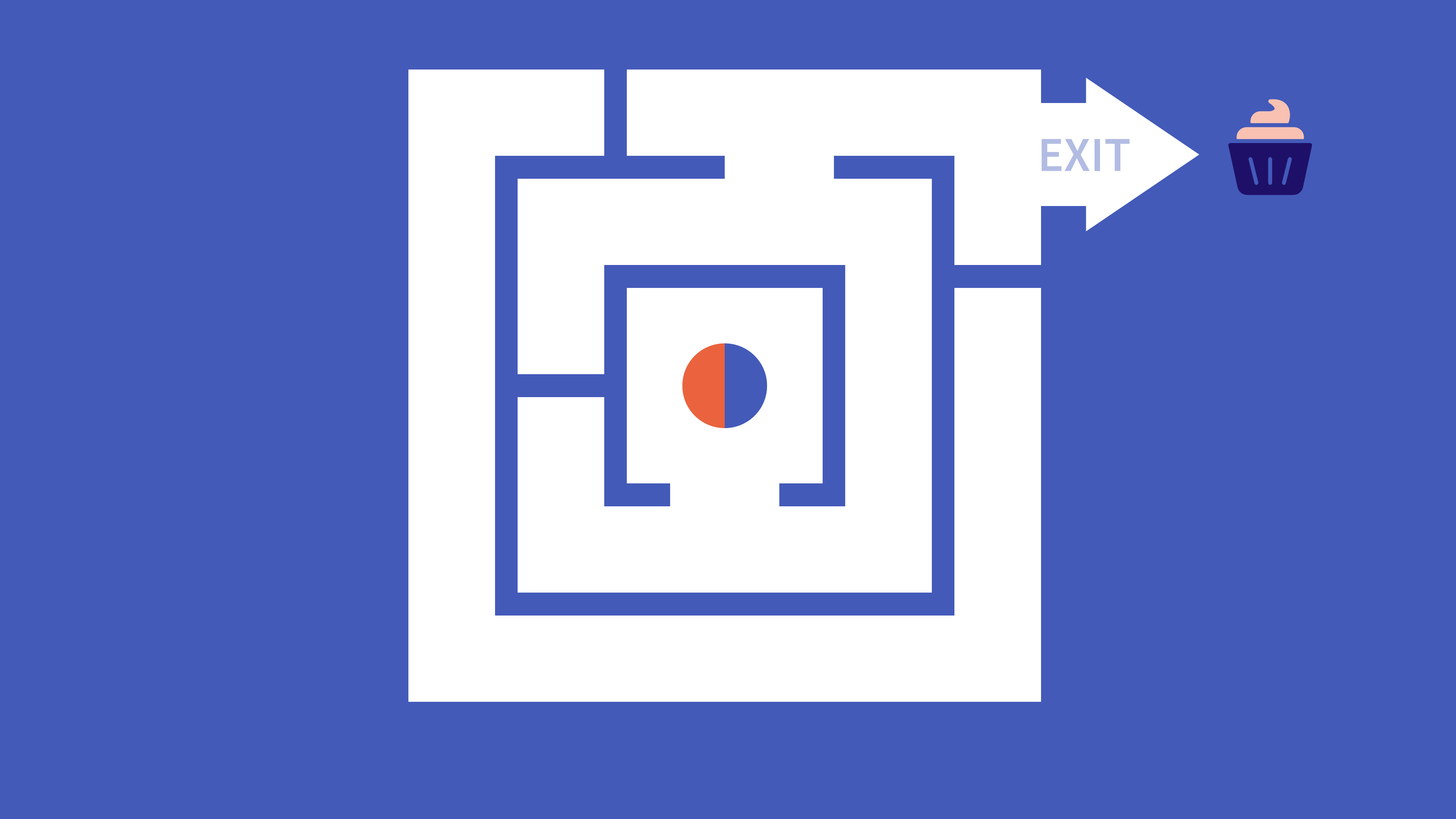}
    \caption{Video still of the maze in the Ant Colony Maze model; one path leads to the exit.}
    \label{fig:screenshot1}
\end{subfigure}
\hfill
\begin{subfigure}{.47\textwidth}
  \centering
\includegraphics[width=\textwidth]{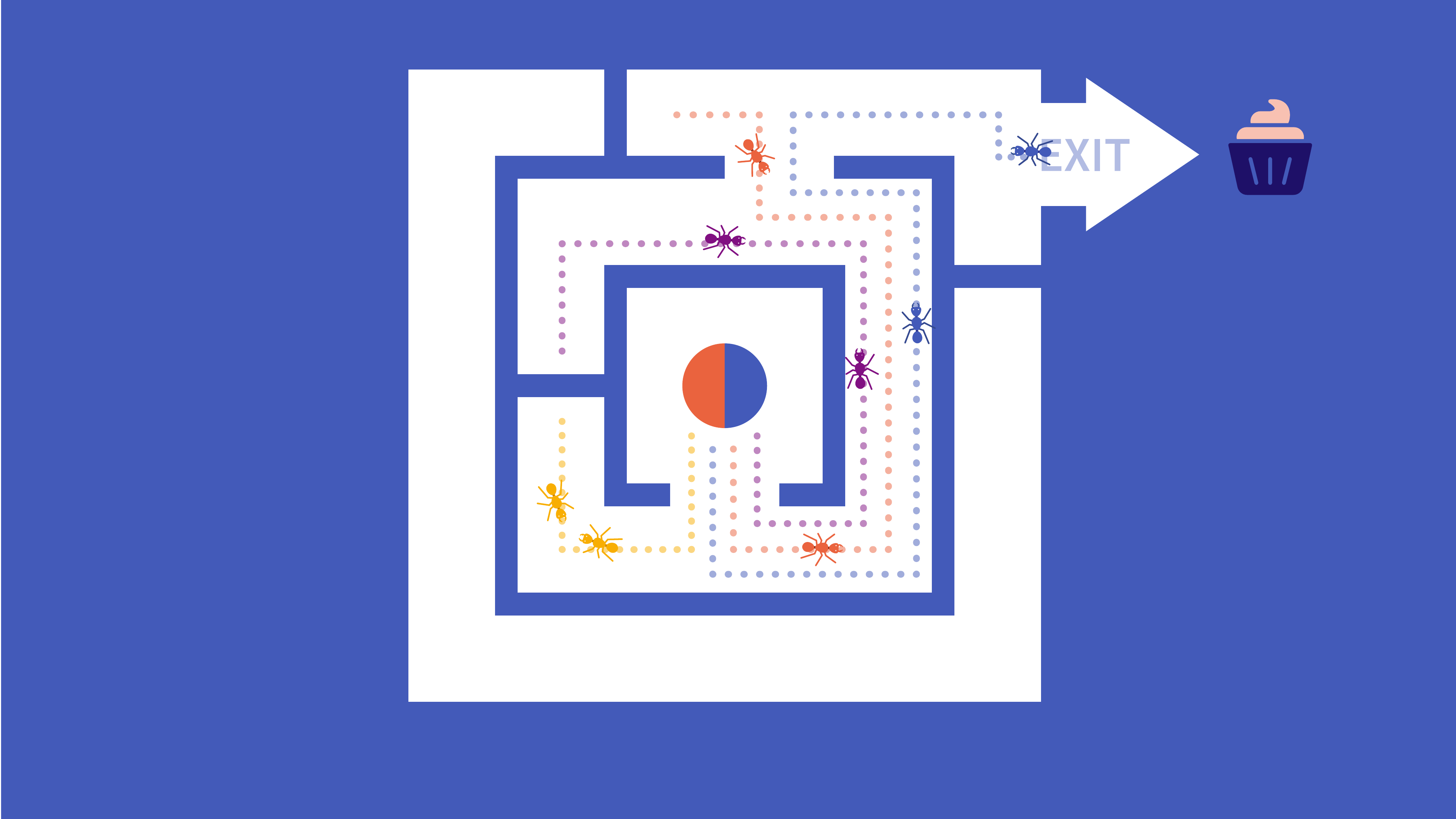}
    \caption{Video still after step 1, after the ants have spread out evenly over the maze.\\\mbox{} }
    \label{fig:screenshot2}
\end{subfigure}
\begin{subfigure}{.47\textwidth}
  \centering
\includegraphics[width=\textwidth]{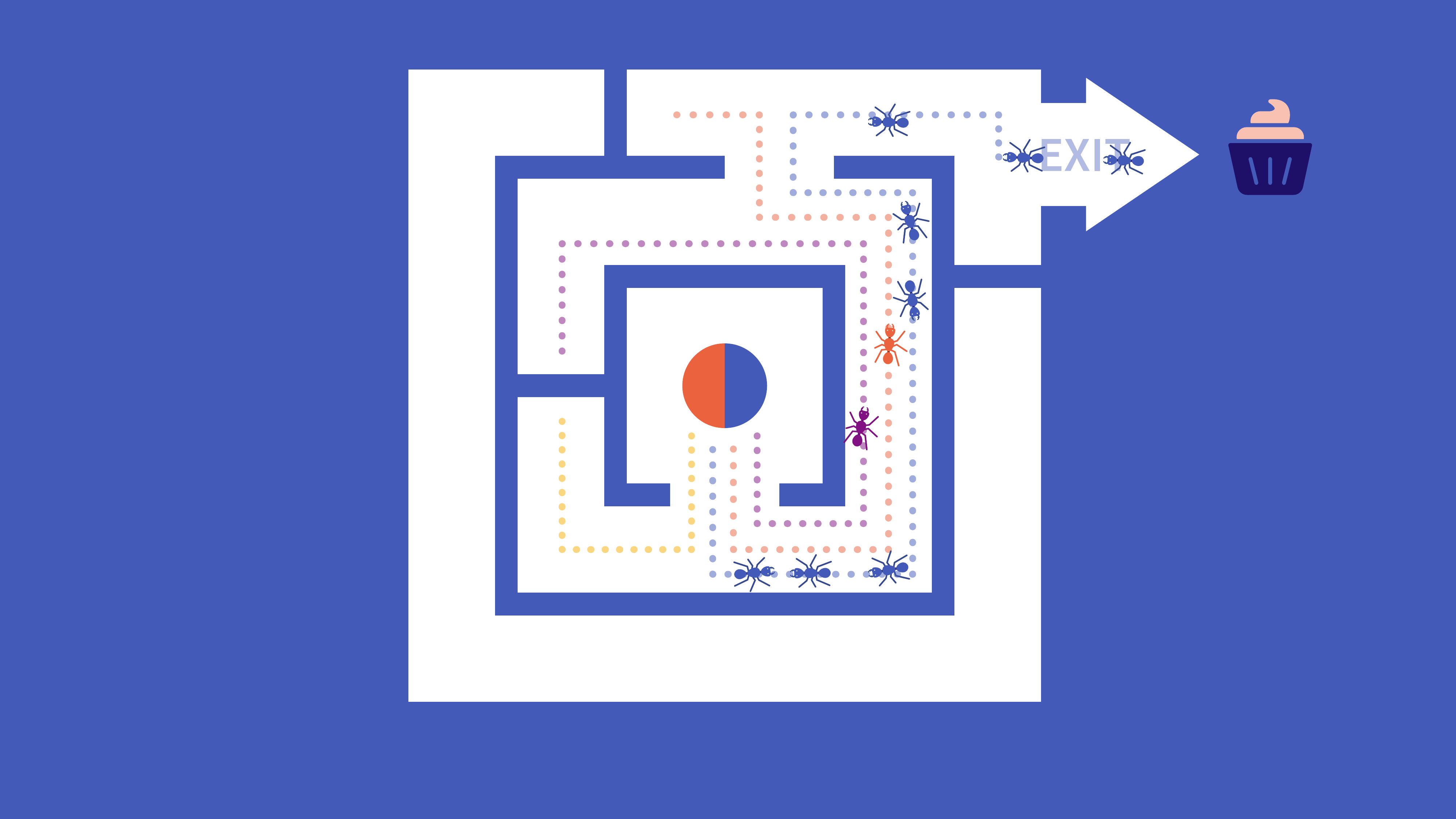}
    \caption{Video still after step 2, when almost all ants can be found on the path to the food.  }
    \label{fig:screenshot4}
\end{subfigure}
\hfill
\begin{subfigure}{.47\textwidth}
  \centering
\includegraphics[width=\textwidth]{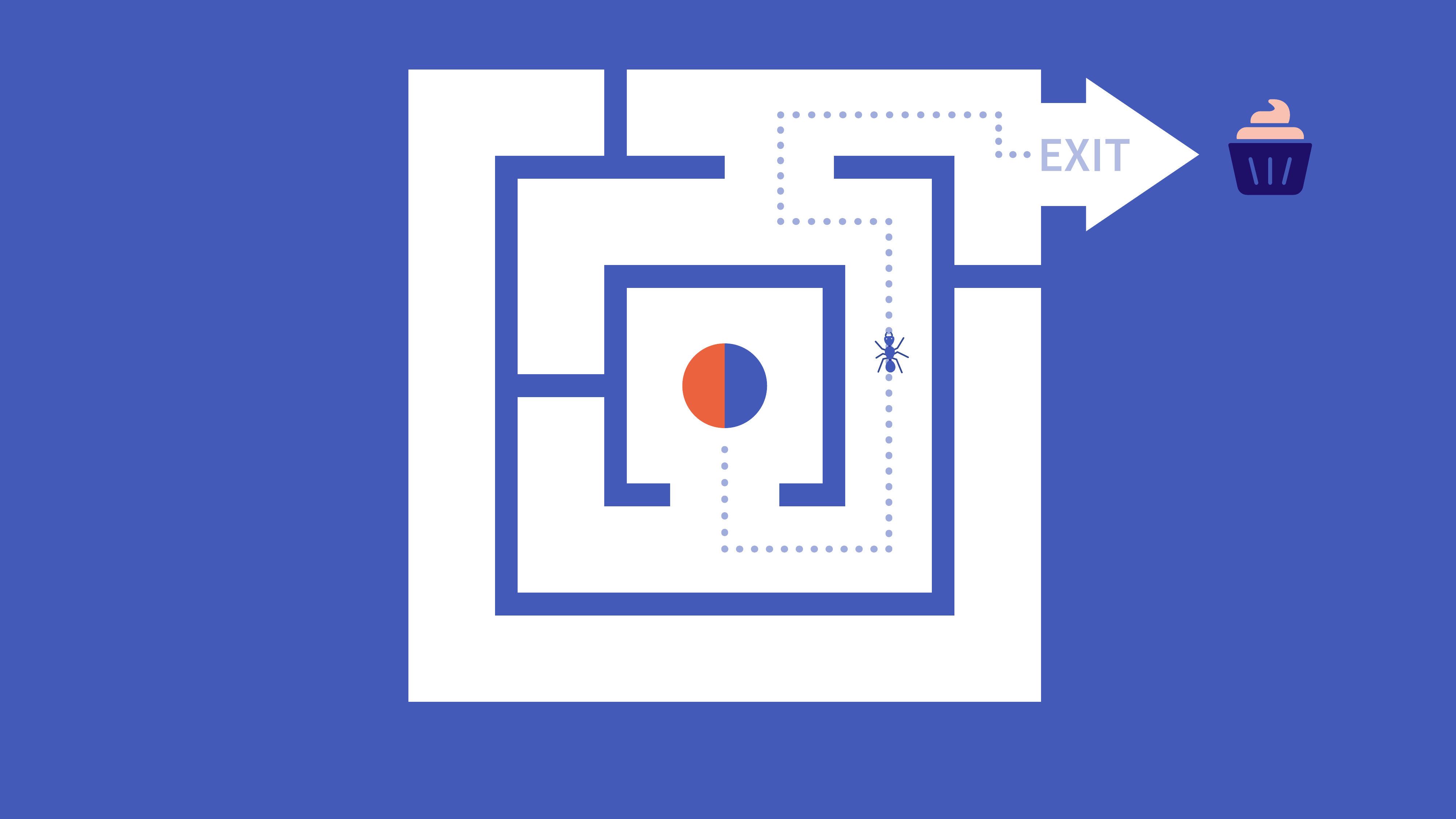}
    \caption{Video still after the measurement in step 3, when one ant is found on the path to the food.}
    \label{fig:screenshot5}
\end{subfigure}
\caption{Video stills showing the different steps of the Ant Colony Maze model.}
\label{fig:qcfd_results_8192}
\end{figure}

\noindent The objective for the ants, of course, is to find food and bring as much as possible back to the ant hill. 
In the first step the ants spread out evenly over the maze as shown in Figure \ref{fig:screenshot2}. This can be interpreted as the start of Grover's algorithm where the quantum state is taken from the $\ket{0}$ state into an equal superposition state $\ket{\mathcal{U}}$, which represents the operation given in Equation \eqref{eq:equal superposition}.

At this point in the video we perform a measurement, i.e., we choose a random ant and its corresponding path. Since all ants are distributed randomly over all paths, the chosen ant does with large likelihood not know the correct path. This strikingly highlights that superposition alone is indeed not useful for finding the solution.

In the second step ants that found food will keep traveling on the same `good' path, collecting food for the colony. Ants that tried a `bad' path and did not find food will try different paths until they find one that leads to the food. This represents the redistribution of weight into the correct solution by Grover's algorithm, and at the same time it mimics the phenomenon of ant trails that are known to many people: communicating via pheromones, ants will conglomerate on the path leading to food relatively quickly. Clearly, ant trails are not quantum computing, but can serve as a mental picture.

The way the distribution of the ants changes in each timestep can be interpreted as the way the state of the system $\ket{\mathcal{U}_l}$ changes to $\ket{\mathcal{U}_{l+1}}$ after applying the Grover iterate for the $l+1$-th time. 
Figure \ref{fig:screenshot4} shows the maze at the end of step 2 where almost all the ants have ended up on the right path, this corresponds to the state of the system after applying the Grover iterate $k$ times, as given in Equation \eqref{eq:final superpos}.

Now, in step 3, upon picking by measurement a random ant in the colony, this ant is most likely on the path which ends in food (see Figure \ref{fig:screenshot5}). To bring that back to Grover's search problem, we can interpret picking an ant on a path that leads to food as the probability of a `good' state to be found after measurement on the state given by Equation~\eqref{eq:final superpos}. 

\section{Existing Visualizations of Grover's algorithm}
\label{sec_visual}
\noindent We submit the Ant Colony Maze model as a pedagogical model for explaining Grover’s algorithm in teaching and for introducing the key quantum-mechanical principles of superposition, interference and measurement. For supporting these claims we describe in this section current visualizations of Grover’s algorithm used in papers and textbooks, and highlight in the next section how the Ant Colony Maze model differs from these visualizations and offers improvements.

\subsection{The arrow diagram}
\begin{figure}[H]
\centering
\includegraphics[width=.8\linewidth]{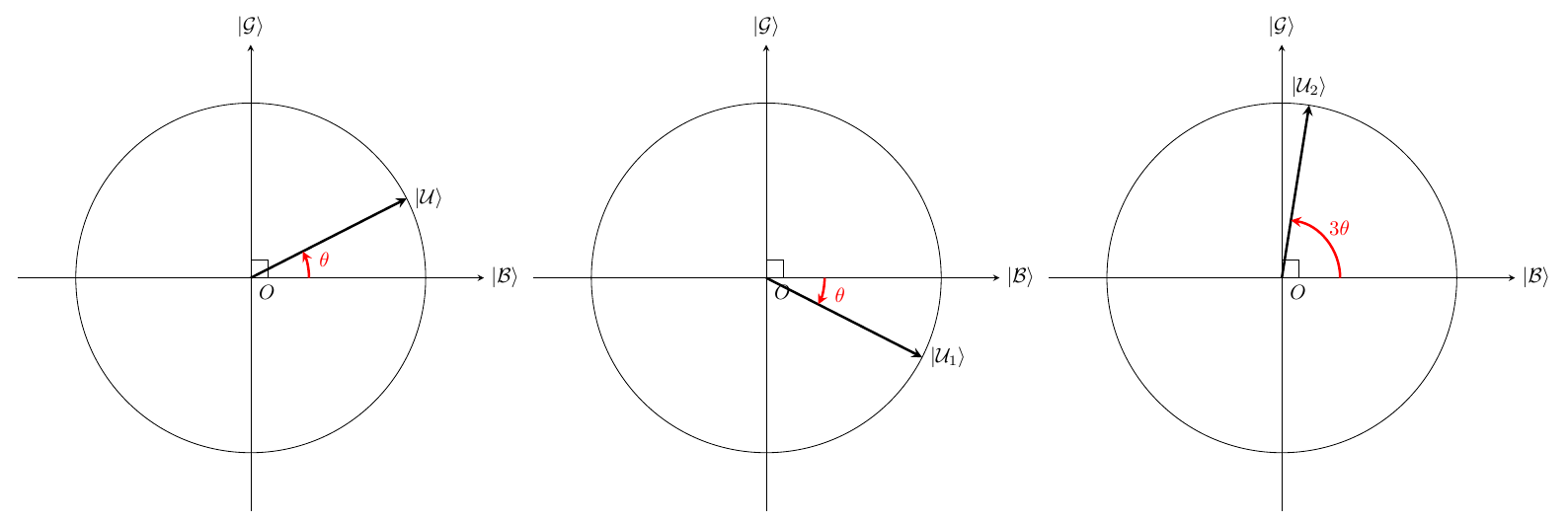}
\caption{Visual representation of the state $\ket{\mathcal{U}}$ and how it is reflected by a Grover iterate to get closer to the state $\ket{\mathcal{G}}$}
\label{fig:arrow_diagram}
\end{figure}

\noindent A peculiar aspect about Grover's algorithm is that the quantum state of the $n$ qubits after each complete Grover iterate can be written as a linear combination of two states $\ket{\mathcal{G}}$ and $\ket{\mathcal{B}}$ with real coefficients, as shown in Equations~\eqref{eq:grover_initial_state}, \eqref{eq:grover_intermediate_state}, and \eqref{eq:final superpos}.
Hence, this quantum state can be immediately represented as a unit vector in two-dimensional space, with the $x$ and $y$-axes representing $\ket{\mathcal{B}}$ and $\ket{\mathcal{G}}$, respectively.
As a consequence, Grover's algorithm can be visualized geometrically \cite{NielsenChuang2010} in the form of an ``arrow diagram". The initial state $\ket{\mathcal{U}}$ has the angle $\theta$ with the $\ket{\mathcal{B}}$-axis as shown in Equation~\eqref{eq:grover_initial_state}.
Hence, the unit vector representing this initial state $\ket{\mathcal{U}}$ is nearly aligned with the $\ket{\mathcal{B}}$-axis, i.e., the angle $\theta$ is small, as the dimension of the solution space of the `good' states is much smaller than $2^n$, as shown in Figure~\ref{fig:arrow_diagram}(a).
The first Grover iterate then corresponds to rotating the unit vector $\ket{\mathcal{U}}$ by an angle $2\theta$ towards the $\ket{\mathcal{G}}$-axis to the unit vector $\ket{\mathcal{U}_{1}}$, as given in Equation~\eqref{eq:grover_intermediate_state}. Hence, the quantum state of the $n$ qubits moves closer to the $\ket{\mathcal{G}}$-axis, as shown in Figure~\ref{fig:arrow_diagram}(b).

The strength of the geometrical visualization is that it represents the mathematical formulas put forward in Section~\ref{subsec_grover}.
In particular, the visualization shows that Grover iterates bring the vector $\ket{\mathcal{U}}$ close to the desired solution $\ket{\mathcal{G}}$ in a finite number of steps.

The arrow diagram clearly serves towards a mathematically inclined reader rather than a broader audience.
Additionally, this visualization has the disadvantage of hiding the typically vastly different dimension of the solution space compared to the complete Hilbert space.
Specifically, the arrow diagram represents the solution space as a 1D subspace in a 2D space. 
Hence, superposition is represented in an abstract way with the unit vector having components in $\ket{\mathcal{G}}$ and $\ket{\mathcal{B}}$, but the fact that the initial state is a \emph{uniform} superposition of all states is ill-represented.
The arrow diagram gives a faithful representation of the Grover iteration mathematics, but the concept of interference is not clearly represented. 
In fact, rotating the arrow seems to rather emphasize a classical mechanism, like turning a knob.
Finally, measurement can be represented as projection on the $x$ and $y$-axis.
However, it again is impeded by the misrepresentation of dimensionality.

\subsection{The bar graph}
\noindent A second commonly employed visualization uses a bar graph to show the (real) amplitude of every quantum state of the $n$ qubits in the Hilbert space.
This was used in the original paper by Grover \cite{Grover1996}, but also for example on the Qiskit website \cite{qiskit_website} alongside the arrow diagram.

\begin{figure}
\includegraphics[width=.8\linewidth]{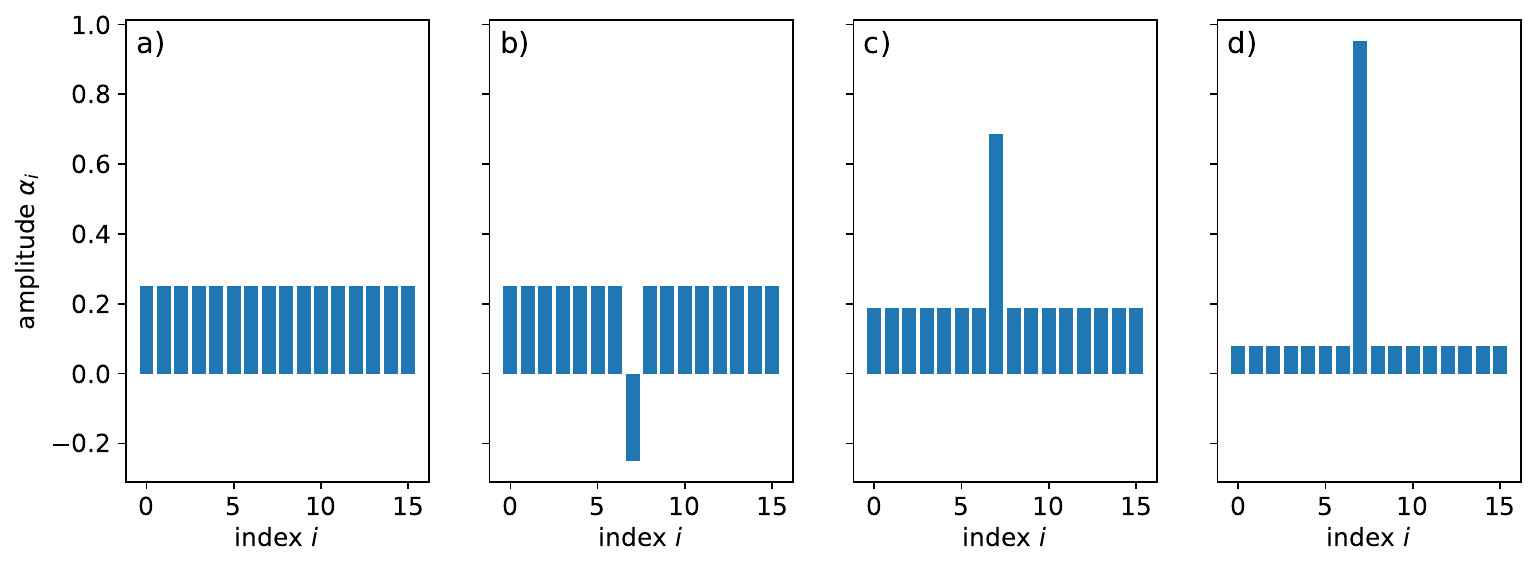}
\caption{A bar diagram represents the real amplitudes of the wave function of (a) the initial state, (b) after application of the oracle, (c) after one Grover iterate, and (d) after two Grover iterates. There is only one solution in this example.}\label{fig:bar_graph}
\end{figure}

The initial state $\ket{\mathcal{U}}$ is now represented by $2^n$ bars of equal height, representing the uniform superposition of all states (Figure~\ref{fig:bar_graph}(a)). 
The action of the oracle function flips the bars corresponding to the solution (Figure~\ref{fig:bar_graph}(b)), and the amplitude of the solution increases with number of Grover iterates (Figures~\ref{fig:bar_graph}(c) and (d)).
In fact, a Grover iterate can be understood as a ``reflection around the average" of the amplitudes.\cite{Grover1996}

The bar graph thus illustrates the concept of superposition well, by showing the equal amplitudes of all quantum states.
It also shows the result of interference -- the amplitude of the solutions grows as the amplitudes of all other states decrease -- though arguably it does not show the process of interference itself.

A disadvantage of the bar graph is the fact that it is a graph and thus not aimed at a broader audience. It is rather a direct representation of the numerical values of quantum mechanical amplitudes.
As such, it is faithful to the underlying mathematics, but also requires an understanding of the mathematics for interpretation.
This particularly impedes the illustrative power of the bar graph to visualize measurement: the knowledge of quantum mechanical probabilities is necessary to infer the principles of measurement from the graph.

\subsection{The Ball Maze model}
\label{sec_ball_model}
\noindent Both the arrow diagram and the bar graph are abstract visualizations: they are agnostic to the particular search problem and do not aim to use a particular example for illustration purposes.
This inherently limits these visualizations to specialists.
A frequently used model to illustrate quantum computing for a non-specialist audience is the maze model, for example used in.\cite{maze_qutech, maze_fermilab, ignite_keynote, ottawa_announcement}
Finding the exit of a maze is a search problem, and can thus be used -- as we also did in the Ant Colony Maze model -- to illustrate Grover's algorithm quite graphically by means of videos of finding paths in the maze.

\begin{figure}[h]
    \centering
    \includegraphics[width = .5\linewidth]{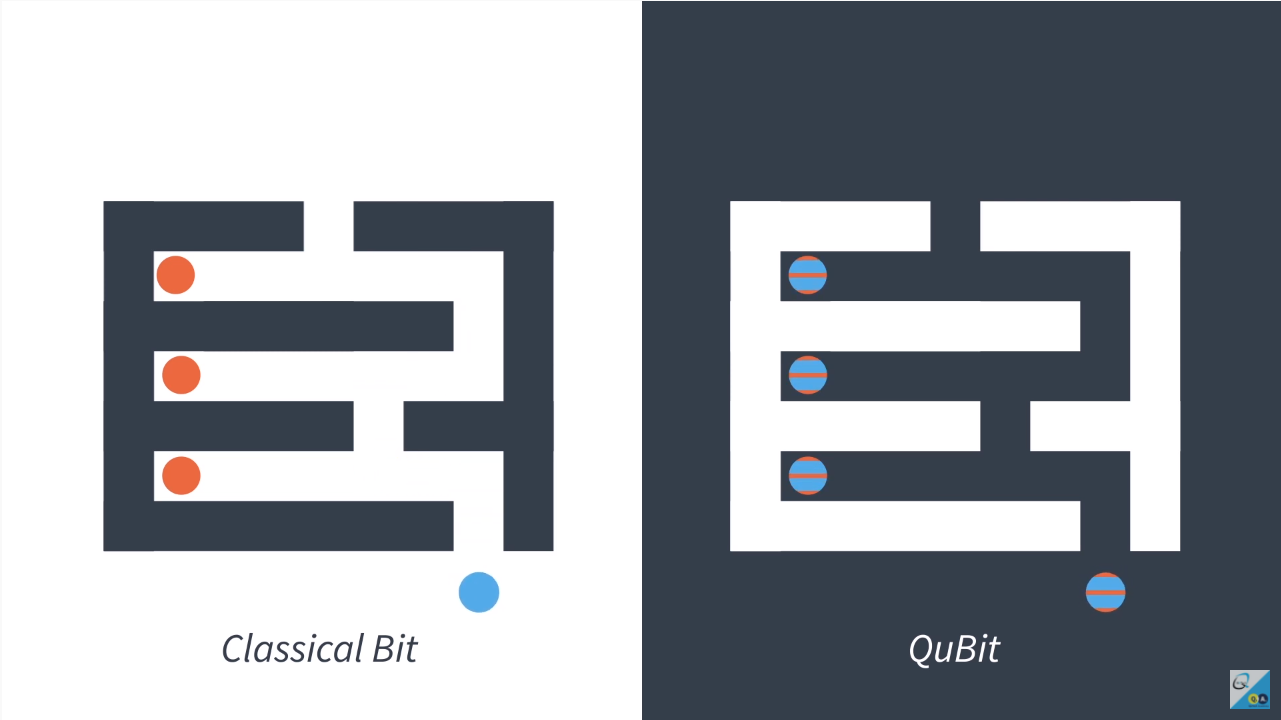}
    \caption{Still from the Ball Maze model in the video from QuTech.}
    \label{fig:ball_model}
\end{figure}

We particularly focus on the maze example used by QuTech Academy \cite{maze_qutech} as a typical representative of this class of visualization, and will refer to it as the \emph{Ball Maze model}.
In this model a ball needs to find the exit of the maze. The classical computer finds the exit by letting the ball sequentially try out each path until the correct one is discovered. The quantum computer has another way of working: each time an intersection is reached the ball splits in a superposition, and each part in that ball superposition then tries a separate pathway after the intersection. In the video it is made certain that the viewer does not have the misconception that the ball itself searches all paths. The ball becomes a superposition and that superposition is then used to find the right path to the exit (see, e.g., Figure~\ref{fig:ball_model}).

The Ball Maze model captures the key principle of superposition in quantum computing in a beautiful and for the student understandable way: the ball represents that the quantum state of the quantum computer evolves to a superposition of the ball, with the parts in the ball superposition rolling along the different paths. Also, by adding an equal color scheme for each ball part in the superposition the principle is made clearer even further. For quantum educated people the wavelike behaviour of quantum states is visible in the color scheme.

The second key principle of quantum interference is not covered by the Ball Maze model. The different balls all remain present in the model, including the ones that hit a dead end in the maze and thus correspond to solutions to the maze that should fade out as balls that have a close to zero probability of being found in the maze when the search has ended. 

Lastly, the third key principle of measurement is also not represented in the Ball Maze model. At the end of the search, when one ball has found the exit, the other less fortunate balls are still visible suggesting that at the end all parts of the superposition still exist. 

\section{Comparison of our model to the Ball Maze model}\label{sec_comparison}
The Ball Maze model is most similar to ours both in ideas and audience. However, our model differs in some essential aspects, making it more powerful, which we highlight here.

Similar to the Ball Maze model, our Ant Colony Maze model captures the idea of superposition by spreading the ant colony over the different paths in a maze. Our model, however, exceeds the Ball Maze model by incorporating how the ``weights'' of the different basis states in the superposition can change over time. By letting the amount of ants walking on a specific path change each timestep, students gain the insight that a superposition must be altered by certain operations to converge towards a final solution. 

This concept ties in with how we explain the measurement principle. By changing the ``weight'' of the superposition over time before the final measurement, we give students insight in how the sought after solution can actually be recovered using a quantum computer. Our model portrays that measurement is a statistical process, and that the likelihood of finding the right state depends on the distribution of the ants over the paths. This is also highlighted in the video by measuring at two different points in time and showing how the (probability of the) outcomes differ. This is a substantial improvement over the Ball Maze model; in that model it appears that just creating an equal superposition of states is enough for correctly identifying the sought after state, whereas we show that the altering of the initial superposition is necessary to converge towards a state where finding the correct state upon measurement is likely.

The principle of interference is, however, a bit harder to capture directly due to the pure mathematical nature of the concept. In our model, unlike the Maze Ball model, we do capture the result of interference (by altering the amount of ants on the paths) but not so much the process itself, similar to the bar graph and arrow diagram models. 

Therefore our model is able to capture two out of the three key principles of quantum computing very well, whilst being able to give insight in the result of the third, while also being a model that does not rely on the purely mathematical visualization of the processes but instead speaks to a broader audience by translation the mathematics of quantum computing to an ``everyday'' example. 

\section{Conclusion}
\label{sec_conclusions}
\noindent 
In this paper we have described a new pedagogical model for explaining quantum computing with Grover's search algorithm, and called it the Ant Colony Maze model. In this model, which comes with a video, finding the exit of a maze represents the search problem to be solved. The strategy by which a colony of ants finds that exit represents Grover's algorithm.

Our motivation to develop this pedagogical model is the need to educate increasingly larger numbers of physicists, computer scientists, and engineers for developing quantum technologies. Grover’s algorithm is central to quantum computing, and for making it understandable it can be divided in three steps that each make use of one of three key principles of quantum mechanics: superposition, interference, and the collapse of quantum states at measurements. The Ant Colony Maze model represents the three steps and thus not only explains to students how the Grover algorithm works, but also introduces these quantum principles. We therefore submit the Ant Colony Maze model as a valuable contribution to education in quantum technologies.

The Ant Colony Maze model builds on a the Ball Maze model, and improves on it since the latter model represents only the principle of superposition explicitly, while the Ant Colony Maze model represents all three quantum mechanical principles. There still is however room for improvement suggesting further research. The principle of interference is hard to capture due to the pure mathematical nature of the concept. In our model its effect is captured by decreases in the number of ants that explores dead-end paths in the maze, yet it does not visualize interference directly by some sort of mechanism that creates that decrease.

An additional advantage of the Ant Colony Maze model is that it can be used to explain to students the differences between computations with quantum computers and with classical computers. The video introduces this difference by representing a search with a classical computer as a search by a single ladybug that explores all paths sequentially. Finally, by its graphic nature, the model may be taken as a means for giving laypersons an informed feel of how quantum computing works, thus also contributing to an understanding of quantum technologies in society at large. The Ant Colony Maze contributes in this way to a development that is already set into motion by the emergence of quantum technologies: counter to Feynman's deposition, understanding the quantum realm will become a privilege of the many.

\newpage

\section*{Author Declarations}
\noindent The authors have no conflicts to disclose.

\newpage

\printbibliography

\end{document}